\documentclass[aps,prl,twocolumn,groupedaddress,showkeys]{revtex4-1}

\usepackage{graphicx}
\usepackage{bm}
\usepackage{amssymb,amsfonts}
\usepackage{amsmath}
\usepackage{soul}
\usepackage{graphicx}
\usepackage[usenames,dvipsnames]{xcolor}
\graphicspath{{Figures/}}

\begin{document}
\title{Dark Matter Spin-Spin Interaction through the Pseudo-Scalar Vacuum Field}%

\author{Alexander I Nesterov}
  \email{nesterov@cencar.udg.mx}
\affiliation{Departamento de F{\'\i}sica, CUCEI, Universidad de Guadalajara,
 CP 44420, Jalisco, M\'exico}

\author{Gennady P  Berman}
 \email{gpb@lanl.gov}
\affiliation{Los Alamos National Laboratory, Theoretical Division, Los Alamos, NM 87545, USA}

\author{Vladimir I Tsifrinovich}
\email{vtsifrin@nyu.edu}
 \affiliation{Department of Applied Physics, NYU Tandon School of Engineering, Brooklyn, NY 
 11201, USA}

\author{Xidi Wang}
 \email{xidiwang@ucsd.edu}
 \affiliation{Department of Chemistry ${\&}$ Biochemistry, University of California San Diego, La Jolla CA 92093, USA}

 \author{Marco Merkli}
   \email{merkli@mun.ca}
 \affiliation{Department of Mathematics and Statistics, Memorial University of 
 Newfoundland, St. John's, NL, Canada A1C 5S7}

\date{\today} 

\begin{abstract}
We suggest that the pseudo-scalar vacuum field (PSV) in the dark matter (DM) sector of the 
Universe may be as important as the electromagnetic vacuum field in the baryonic sector. In 
particular, the spin-spin interaction between the DM fermions, mediated by PSV, may represent 
the strongest interaction between the DM fermions due to the absence of the electric charge 
and the magnetic dipole moment. Based on this assumption, we consider the influence of the 
spin-spin interaction, mediated by PSV, on the spin precession of the DM fermions  (e. g. 
neutralino).  In the secular approximation, we obtain the exact expression describing the 
frequency of the precession and estimate the decoherence rate.  
\end{abstract}




\maketitle

\section{Introduction}

{It is well-known that the pseudo-scalar vacuum (PSV) field (e.g. axionic field) interacts with 
the spins of the baryonic matter (see, for example, \cite{Flambaum,Flambaum1}). As a result, 
PSV mediates the spin-spin interactions in the baryonic matter  (see, for example, \cite{Daido}). 
 Certainly, interactions between the PSV and baryonic matter represent a tiny correction to the 
 electromagnetic interactions. We suggest that in the dark sector of the Universe the situation 
 may be opposite: due to the absence of the electric charge and the  magnetic moment, the 
 interaction between the dark matter (DM) fermionic spins (e.g. neutralino spins) and PSV may 
 manifest the leading interaction in DM. (We note here that neutralino remains one of the 
 main candidates for DM \cite{Shafi,Hamaguchi,Bramante,Ypshimatsu,Fukuda}. However, there are many other fermionic candidates as well - see, for example, \cite{Ruffini1,Ruffini2}.)
 In particular, the DM fermionic spins could produce permanent fields similar to the magnetic 
 fields in the baryonic sector. 
 
 Based on this assumption we analyze precession of the two DM 
 fermionic spins, coupled through PSV, in a non-uniform external field produced by the other 
 DM fermions. We assume that the two DM spins are located at the points with the position 
 vectors, ${\mathbf r}_1$ and ${\mathbf r}_2$. The frequencies of the spin precession in the 
 external axion field are given by $\varepsilon_{1}$ and $\varepsilon_{2}$. In order to obtain 
 the analytical expressions describing the spin precession, we assume that $\varepsilon_1$,
 $\varepsilon_2$ and $|\varepsilon_1 - \varepsilon_2|$ are  larger than the interaction between 
 the axion field and spins. This allows us to use the secular approximation well-known in 
 magnetic resonance (see,  for example, \cite{Schum}). 
 
 Note, that we limited ourselves by considering the consequences of the 
    interaction of the DM fermionic spins through the PSV. So, we do not 
    consider  many other very important and widely discussed problems such as chiral anomaly 
    of fermions \cite{Mielke}, the issues related to ultra-light axions \cite{Calabrese,Berman1}, 
    many existing protocols for axion and neutralino detection \cite{Goryachev,Aaboud,BGHD}, etc.
 }
 
In this paper, we analyze precession of the two DM fermionic spins $1/2$, coupled through 
PSV, in a non-uniform  external axion field ($\sim\nabla\varphi({\mathbf r})$) produced by the 
other DM fermions or by axion-originated topological defects.
The frequencies of the spin precession in the external axion field are given by:  
$\varepsilon_{1,2}=2\lambda|\nabla\varphi({\mathbf r}_{1,2})|/\hbar$. It is important to note 
that: (i) the frequencies, $\varepsilon_{1,2}$ are analogous to the well-known nuclear magnetic 
resonance (NMR) or electron paramagnetic resonance (EPR) frequencies of precession of single 
nuclear or electron spin in the permanent magnetic field, (ii) these frequencies are proportional 
to the first order of the coupling constant, $\lambda$, and (iii) the search of axions, discussed  
in \cite{Flambaum,Flambaum1}, is based, in particular, on detection of precession 
frequencies of electron or nuclear spins in axion field. 

In our paper, we are interested in a different effect, namely, in the interaction of two DM 
spins through the PSV. We demonstrate that in this case, one spin rotates around 
the other non-rotating spin due to their interaction through the PSV.  So, this 
effect is (i) of the second order in $\lambda$, (ii) analogous to the well-known  Lamb shift of 
energy levels in the vacuum electromagnetic field, and (iii)  has no direct relation to the  
$\varepsilon_{1,2}$ spin precession.

The interaction of the two spins with the permanent field and PSV can be written as (see, for 
example, \cite{Flambaum,Flambaum1}),
\begin{align}
	{ H}_{int} =  \lambda \sum_{\alpha=1,2}{\nabla \varphi({\mathbf r}_{\alpha})} \cdot  {\boldsymbol 
	\sigma}^\alpha.
	\label{EqH}
\end{align}
Here $\varphi$, as above, is the PSV, $\lambda$ is the coupling constant, and 
${\boldsymbol \sigma}^\alpha \equiv (\sigma_x^\alpha,\sigma_y^\alpha,\sigma_z^\alpha)$ is 
the vector built by Pauli matrices of spin $\alpha$, located at $\mathbf r_\alpha$. 

Through this paper, we will use the natural units convention, $\hbar =c =1$. 

\section{Precession of the spins} 	
	
Our total Hamiltonian, $H$, describes, in the secular approximation, the two DM fermionic spins interacting with the external field, oriented in the $z$-direction (direction of the $\nabla\varphi$), and the PSV: 
\begin{align}
	{H} = & \frac{1}{2}\sum_{\alpha=1,2} {\varepsilon_\alpha}\sigma^\alpha_{ z }+ 
	\sum_{\mathbf k}\omega_k \hat 
	a^{\dagger}_{\mathbf k} \hat a_{\mathbf k} \nonumber \\
	&+ \frac{i\lambda }{\sqrt{V}} \sum_{\alpha=1,2} \sigma^\alpha_z
	\sum_{\mathbf k} \frac{{k_z} }{\sqrt{2\omega_k}} \big (\hat a_{\mathbf k} e ^{i\mathbf k\cdot \mathbf r _\alpha}- \hat 
	a^{\dagger}_{\mathbf k} e ^{-i\mathbf k\cdot \mathbf r_\alpha} \big).
\label{H0}
\end{align}
Here $\varepsilon_\alpha$ is the transition frequency of spin $\alpha$ in the external permanent field, $\omega_k=\sqrt{k^2+m^2}$ is the frequency of the field mode with wave number $\mathbf k$, where $m$ is the mass of a DM boson, and $V$ is the quantization volume. 

We will consider the initial state, $|\psi \rangle$, of the spin-PSV system  as the tensor product of the spin $|\psi_s \rangle$ and the PSV $|\psi_V \rangle$ states: $|\psi\rangle =|\psi_s \rangle \otimes |\psi_V \rangle$. Below, we use the $\sigma_z$-representation for the single spin states: vectors $|0_\alpha \rangle$ and $|1_\alpha \rangle$ ($\alpha=1,2$) denote the spin pointing in the negative and positive $z$-direction, respectively. For two spins, we use the basis:  $|0\rangle\equiv|0_20_1\rangle$,~$|1\rangle\equiv|0_21_1\rangle$,~$|2\rangle\equiv|1_20_1\rangle$,~$|3\rangle\equiv|1_21_1\rangle$. Then, the initial spin wave function is, $|\psi_s\rangle=\sum_{i=0}^3C_i|i\rangle$, ($\sum_{i=0}^3|C_i|^2=1$). The corresponding initial spin density matrix is, 
\begin{align}
\rho_s(0)=\sum_{i,j=0}^3\rho_{ij}(0)|i\rangle\langle j|,~(\rho_{ij}(0)=C_iC_j^*).      
\label{R}
\end{align}
The concrete values of the amplitudes, $C_i$, are not important for us. 
In the interaction representation, the evolution operator of the spin-PSV system can be written as \cite{BKT},
\begin{widetext}
\begin{align}
U(t) =\hat T \exp \Big(- \frac{i}{\hbar }\int_0^t dt' H_{int}(t')\Big ) = \exp \Big(\frac{i}{2}\sum_{\alpha,\beta=1,2} \sigma^\alpha_z \sigma^\beta_z \nu_{\alpha \beta}(t) \Big )
 \exp \Big(\sum_{\alpha=1,2} \sigma^\alpha_z\sum_{k} \big (e ^{-i\mathbf k\cdot \mathbf r_\alpha} \xi_k(t) \hat a^\dagger_{\mathbf k}  -  
e ^{i\mathbf k\cdot \mathbf r_\alpha}  {\xi}^\ast_k(t) \hat a_{\mathbf k} \big)\Big ),
\label{U1}
\end{align}
\end{widetext}
where 
\begin{align}
\xi_k (t) = \frac{\lambda k_z(e^{i \omega_k t}- 
1)}{\sqrt{2V}\omega_k^{3/2}},
\label{xi}
\end{align}
and
\begin{align}
\nu_{\alpha \beta} =i\sum_{\mathbf k} \cos (\mathbf k\cdot \mathbf r_{\alpha\beta}) 
\int^t_{0}dt'\big(  \xi_k(t'){\dot \xi}^\ast_k(t') - 
\xi^\ast_k(t'){\dot \xi}_k(t') \big ).
\label{G12}
\end{align}
Here $\mathbf r_{\alpha\beta} = \mathbf r_\beta - \mathbf r_\alpha $.

Using the evolution operator $U(t)$ of  Eq. \eqref{U1}, one can write the total density matrix as,
\begin{align}
	\rho_{tot} (t) =U(t)\rho_{tot} (0)U^{-1}(t),
	\label{rho}
\end{align}
where $\rho_{tot} (0)=|\psi \rangle \langle \psi |$. Next, we have obtained the 
$4\times 4$ reduced spin density matrix, $\rho_s(t)$, by tracing out the PSV degrees 
of freedom. (The details of computations are given in Appendix A.)

 Using Eqs. (\ref{R}) and (\ref{rho}), we calculate the $x$-component of the total spin:
 \begin{align}
 \bar 
 S_x(t)=&\sum_{\alpha=1,2} Tr(\rho_{s}(t)\sigma_x^{\alpha})=\sum_{\alpha=1,2}\sum_{i=0}^3\langle
  i|\rho_{s}(t)\sigma_x^{\alpha})|i\rangle\nonumber \\
 =&\rho_{01}(t)+\rho_{02}(t)+\rho_{13}(t)+\rho_{23}(t)+c.c.,
 \label{m}
 \end{align}
where $\rho_{ij}(t)$ are the time-dependent components of the reduced spin density 
matrix. One can see that $ \bar S_x(t)$ is associated with the precession of a single spin while the other spin remains in a basis state.  

For definiteness, consider the element $\rho_{01}(t)$  of the reduced spin density matrix. We have derived the following expression:  
\begin{align}
	\rho_{01}(t) = \rho_{01}(0) e^{i\delta (t) - \gamma (t) },
\end{align}
where,
\begin{align}\label{D8}
\delta (t)= &-2\big (\nu_{12}(t) - \sum_{\mathbf k}\sin (\mathbf k\cdot \mathbf r_{12})|\xi_k(t)|^2\big ), \\
	{\gamma(t)}=&2 {\sum_{\mathbf k}|\xi_{k}(t)|^2} .
	\label{D9}
\end{align} 

The functions, $\delta (t)$ and $\gamma (t)$, describe the frequency shift and decoherence of the spin precession due to the indirect interaction between the fermionic spins through the PSV. Substituting the expression \eqref{xi} for $\xi_k(t)$  and changing sums to integrals,  $(1/V )\sum_{\mathbf k} \rightarrow (1/(2\pi)^3)\int d^3k$,  we obtain:
\begin{align}\label{D10e}
	 {\delta(t)}=&-\frac{\lambda^2}{4 \pi^3}\int d^3k\bigg( \frac{k_z^2 \cos (\mathbf k\cdot \mathbf r_{12})(\omega_k t
	 -\sin(\omega_k t))}{\omega_k^3}\nonumber \\
	&-  \frac{2k_z^2 \sin (\mathbf k\cdot \mathbf r_{12})\sin^2(\omega_kt/2)}{\omega_k^3}\bigg ), \\
	 \gamma(t)=&\frac{\lambda^2} {4\pi^3 }\int \frac{d^3k 
	 k_z^2\sin^2(\omega_kt/2)}{\omega_k^3} .
	 \label{D10g}
	 \end{align}

{\em The frequency shift.} -- We computed the 3D integral in Eq. \eqref{D10e}, assuming that $mL\ll 1$, where $L=|{\bf r}_{12}|$, (see Appendix B for details), and obtained the following expression for the phase shift:
\begin{align}
	\delta (t)= \omega_{s}t,~\omega_{s}=\frac{\lambda^2}{\pi}\cdot\frac{3 \cos^2 \theta -1}{r^3_{12}}.
	\label{D11}
\end{align}
Here $\omega_s$ is the frequency shift, $\theta$ is the polar angle of the vector $\mathbf r_{12}$ connecting the two spins. Formula \eqref{D11} is valid for  $t > r_{12}$; for $t < r_{12}$ we have $\delta (t) = 0$. 

One can see that the phase shift is proportional to time. The coefficient  at $t$ is the frequency 
shift, $\omega_s$, which is proportional to $(3 \cos^2 \theta -1)/r^3_{12}$ . This factor is similar to that 
obtained for the magnetic dipole-dipole interaction in the baryonic matter 
\cite{Schum}, in spite of the magnetic moment for DM fermion is zero. In both 
cases the frequency shift changes its sign at the magic angle, $\cos^2\theta=1/3$. Thus, we 
come to the conclusion that the spin-spin fermionic interaction, mediated by PSV, is similar to 
the magnetic dipole-dipole interaction mediated by the vacuum electromagnetic field.\\

{\em Decoherence.} -- The integral in Eq. \eqref{D10g}, describing the decoherence, is 
diverging. This divergence is non-physical. It can be eliminated with proper renormalizations.
 In Appendix A we compute this integral using the dimensional 
regularization. \\
	
{\em Generation of entanglement entropy.} -- By tracing out the pseudo-scalar 
degrees of freedom, we generate the entanglement entropy (EE), 
$E(t)=-Tr(\rho_s(t)\ln\rho_s(t))$, in the spin sub-system. The EE is created 
independently of the initial spin wave function (with $E(0)=0$) disentangled or 
entangled. Namely, in both cases, the trace over the pseudo-scalar vacuum field 
creates entanglement for spin sub-system, at $t>0$ ($E(t) > 0$). 
	
\section{Conclusion}

We suggested that in the dark sector of the Universe the interaction between the PSV of the DM 
bosons and DM fermions may be as important as the electromagnetic interaction in the baryonic 
sector. In particular, the spin-spin interaction between the DM fermions, mediated by PSV, may 
represent the leading interaction in DM.

Based on this assumption, we consider the following situation in the dark sector of the 
Universe 
(below, for definiteness, we will write “neutralinos” and “axions” instead of DM fermions and 
bosons). In the first scenario, a large ensemble of neutralino spins (ENS) produces a strong permanent axionic field. 
This is similar to the permanent magnetic field produced by a large ensemble of electron spins 
of a permanent magnet in the baryonic sector. In the second scenario, the permanent axionic 
field is produced by axion topological defects of soliton or domain wall types 
\cite{Flambaum,Flambaum1}.

Next, we consider the system of two neutralino spins which experience (i) the permanent axionic field produced by the ENS or by axion-generated topological defects and (ii) indirect interaction between themselves through the axion vacuum field. Each neutralino spin precesses in the permanent axionic field like nuclear or electron spin precesses in the magnetic field of the permanent magnet. The frequency of this precession is proportional to the first order in the perturbation constant, $\lambda$, and is a subject of intensive experimental research for axions by using electron or nuclear spins. (See, for example, \cite{Flambaum,Flambaum1}, and references therein.) Note that the neutralino spin precession would generate radiation of the real axions which could be detected experimentally like electron spin precession generates radiation of the real photons detected by the NMR and EPR  techniques. (In this work we do not explore the process of radiation of real axions). 
  
   Our main attention is concentrated on an indirect interaction between the two neutralino spins mediated by the vacuum  axionic field like the magnetic dipole-dipole interaction between the electron spins is 
   mediated by the electromagnetic vacuum. The indirect spin-spin interaction is the only 
   non-gravitational interaction between the neutralinos as these particles do not have an 
   electric charge and the magnetic moment. In this work, we have studied the influence of the 
   indirect interaction between the neutralino spins on the spin precession. In the secular 
   approximation, we have derived exact analytical expressions for the frequency shift and 
   decoherence rate of the neutralino spin precession caused by the interaction mediated by the 
   vacuum axionic field. We demonstrate that this frequency shift is analogous to the Lamb shift of energy levels in electromagnetic vacuum, and it is proportional to the second order in the perturbation constant, $\lambda$. The mechanism of this frequency shift is related to the precession of one spin around the other spin, due to their interaction through the axion vacuum field. This mechanism of spin precession has no direct relation to the spin precession discussed in recent proposals (see \cite{Flambaum,Flambaum1}, and references therein). Certainly, our results are valid not only for neutralinos and axions but 
   for any DM fermions interacting with the PSV of the DM bosons.

\begin{acknowledgments}
The work by G.P.B. was done at Los Alamos National Laboratory  managed by Triad National Security, LLC, for the National Nuclear Security Administration of the U.S. Department of Energy under Contract No. 89233218CNA000001. A.I.N. acknowledges the support by CONACyT (Network Project No. 294625, ``Agujeros Negros y Ondas Gravitatorias'').  M.M. was supported by an National Sciences and Engineering Research Council of Canada (NSERC)  Discovery Grant.
\end{acknowledgments}

\appendix
\section{Supplemental Material}
\subsection{Appendix A}
\subsection{Evolution operator} 

In this section, we derive the elements of the reduced spin density matrix.
First, consider the evolution operator written as,
\begin{align}
	U(t) = \exp \Big(\frac{i}{2}\sum_{\alpha,\beta} \sigma^\alpha_z \sigma^\beta_z \nu_{\alpha \beta}(t) \Big )D(\xi_k(t) ),\,\alpha,\beta =1,2,
\end{align}
where $D(\xi_k ) $ is the  displacement operator:
\begin{align}
D(\xi_k(t) )= \exp \Big(\sum_{\alpha, \mathbf k} \sigma^\alpha_z\big (e ^{-i\mathbf k\cdot \mathbf r_\alpha} \xi_k(t)  \hat a^\dagger_{\mathbf k}  -  
e ^{i\mathbf k\cdot \mathbf r_\alpha}  {\xi}^\ast_k(t)\hat a_{\mathbf k} \big)\Big ).
\label{SU1}
\end{align}
Here,
\begin{align}\label{Axi}
\xi_k (t) = & \frac{\lambda k_z(e^{i \omega_k t}- 
1)}{\sqrt{2V}\omega_k^{3/2}},\\
\nu_{\alpha \beta} =&i\sum_{\mathbf k} \cos (\mathbf k\cdot \mathbf r_{\alpha\beta}) 
\int^t_{0}dt'\big(  \xi_k(t'){\dot \xi}^\ast_k(t') - 
\xi^\ast_k(t'){\dot \xi}_k(t') \big ),
\label{AG12}
\end{align}
and we set, $\mathbf r_{\alpha\beta} = \mathbf r_\beta - \mathbf r_\alpha $.

The  displacement operator can be recast as,
\begin{align}
D(\xi_k ) = e^{-\kappa(t)/2}e^{\sum_{\alpha} \sigma^\alpha_z e ^{-i\mathbf k\cdot \mathbf r_\alpha} \xi_k \hat a^\dagger_{\mathbf k}} e^{-\sum_{\alpha} \sigma^\alpha_z e ^{i\mathbf k\cdot \mathbf r_\alpha}  \xi^{\ast}_k \hat a_k }  ,
\label{SD2}
\end{align}
where $\kappa(t) =\sum_{\alpha,\beta} \mu_{\alpha \beta} (t)\sigma^\alpha_z \sigma^\beta_z$ and $\mu_{\alpha \beta} (t) = e ^{i\mathbf k\cdot \mathbf r_{\alpha \beta}} |\xi_k(t)|^2$. Taking into account that $\hat a_{\mathbf k} |\Psi_V\rangle =0$, we obtain,
\begin{align}
D(\xi_k) |\Psi_V\rangle =e^{-\kappa(t)/2}\sum_{m}\frac{(\sum_{\alpha} \sigma^\alpha_z e ^{-i\mathbf k\cdot \mathbf r_\alpha} \xi_k)^{m}}{{m!} }  (\hat a^\dagger_{\mathbf k})^{m}|\Psi_{V}\rangle.
	\label{SAD1}
\end{align}
Next, using the relation, $(\hat a^\dagger_{\mathbf k})^m  |\Psi_V\rangle   =\sqrt{m !} \,|m_{\mathbf k}\rangle $, we find,
\begin{align}
D(\xi_k) |\Psi_V\rangle =e^{-\kappa(t)/2}\sum_{m}\frac{(\sum_{\alpha} \sigma^\alpha_z e ^{-i\mathbf k\cdot \mathbf r_\alpha} \xi_k)^{m}}{\sqrt{m!} }   |m_{\mathbf k}\rangle. 
\label{AD2}
\end{align}

Before proceeding further, it is convenient to introduce a new auxiliary displacement operator:
\begin{align}
D(\xi^a_k ) =e^{\xi^a_k\hat  a^\dagger_{\mathbf k} - \xi^{a\ast}_k \hat a_{\mathbf k} } = e^{-|\xi^a_k|^2/2}e^{ \xi^a_k \hat a^\dagger_{\mathbf k}  } e^{-  \xi^{a\ast}_k \hat a_{\mathbf k}}  ,
\label{SD2a}
\end{align}
where $\xi^a_k = \kappa_a \xi_k$ ($a=1,2$), with $\kappa_1 = \sum_{\alpha} e ^{-i\mathbf k\cdot \mathbf r_\alpha} $ and $\kappa_2 = \sum_{\alpha} (-1)^\alpha e ^{-i\mathbf k\cdot \mathbf r_\alpha} $.

Now, applying the evolution operator to the basis states, we obtain,
\begin{align}\label{AD1a}
U|0 \rangle\otimes |\Psi_V\rangle = &e^{i(\nu + \nu_{12} ) 
}|0 \rangle\otimes\prod_k D\big (-\xi^1_k\big )|\Psi_V\rangle, \\
U| 1\rangle\otimes |\Psi_V\rangle =& e^{i(\nu - \nu_{12} ) }
| 1\rangle\otimes \prod_k D\big(  
-\xi^2_k\big)|\Psi_V\rangle , \\
U|2 \rangle\otimes |\Psi_V\rangle = &e^{i(\nu - \nu_{12} ) 
}|2 \rangle\otimes\prod_k D\big (\xi^2_k\big )|\Psi_V\rangle, \\
U| 3\rangle\otimes |\Psi_V\rangle =& e^{i(\nu + \nu_{12} ) }
| 3\rangle\otimes \prod_k D\big(  
\xi^1_k\big)|\Psi_V\rangle,
\label{AD1b}
\end{align}
where $\nu =(1/2)(\nu_{11} + \nu_{22})$.

The matrix elements of the reduced density matrix are given by,
\begin{align}
	\rho_{ij}(t) = \langle i| {\rm Tr}_R U(t)\varrho(0) U^{-1}(t)|j \rangle, \quad 
	i,j=0,1,2,3,
	\label{D4a}
\end{align}
where $\varrho(0) = \rho_s(0)\otimes|\Psi_V\rangle\langle\Psi_V|$.
We start with calculation of the matrix element $\rho_{01} (t)$.  Using \eqref{AD1a} --  \eqref{AD1b} in \eqref{D4a}, we obtain,
\begin{align}
	\rho_{01} (t)= e^{-2i\nu_{12}(\tau)  } e^{-\mu(t)}\rho_{01} (0),
	\label{SD5}
\end{align}
where
\begin{align}
	\mu=\frac{1}{2}\sum_{\alpha,\beta} \big(1+ (-1)^{\alpha - \beta} \big)\mu_{\alpha \beta}- \xi^{1\ast}_k\xi^2_k.
\end{align}
Substituting $\xi^a_k =\kappa_a\xi_k $, into expression for $\mu$, after some algebra we obtain,
\begin{align}
	{\mu}=&2\sum_{\mathbf k}|\xi_{k}(t)|^2 
	+ 2i\sum_{\mathbf k}|\xi_{k}(t)|^2\sin(\mathbf k \cdot \mathbf r_{12}). 
	\label{SD9}
\end{align} 
Now, taking into account all these relations, we get, 
\begin{align}
	\rho_{01}(t)= e^{i\delta(t) } e^{-\gamma(t)}\rho_{01} (0), 
	\label{AD10a}
\end{align}
where,
\begin{align}\label{AD11a}
\delta(t)= &-2\big (\nu_{12}(t) + \sum_{\mathbf k}\sin (\mathbf k\cdot \mathbf r_{12})|\xi_k(t)|^2\big ), \\
	{\gamma(t)}=&2 {\sum_{\mathbf k}|\xi_{k}(t)|^2}.
	\label{AD11}
\end{align} 
Similar computation of the other  elements of the reduced density matrix yields our final result:
\begin{align}
\rho_{ii} (t)= &\rho_{ii} (0), \,\, i=0,1,2,3 ,\\
\rho_{01} (t)= &e^{i\delta(t) } e^{-\gamma(t)}\rho_{02} (0), \\
\rho_{02} (t)= &e^{i\delta(t) } e^{-\gamma(t)}\rho_{02} (0), \\
\rho_{03} (t)=&e^{- \gamma_1 (t)}\rho_{03} (0),\\
\rho_{12} (t)=&e^{- \gamma_2 (t)}\rho_{12} (0), \\
\rho_{13} (t)= &e^{-i\delta(t) } e^{-\gamma(t)}\rho_{13} (0), \\
\rho_{23} (t)= &e^{-i\delta(t) } e^{-\gamma(t)}\rho_{23} (0).
\end{align}
Here,
 \begin{align}
 \delta(t)= &-2\nu_{12}(t) +2 \sum_{\mathbf k}\sin (\mathbf k\cdot \mathbf r_{12})|\xi_k(t)|^2, \\
 	 {\gamma(t)}=&2 {\sum_{\mathbf k}|\xi_{k}(t)|^2} , \\
 	 \gamma_1 (t) = &8 \sum_{\mathbf k} \cos^2\Big (\frac{\mathbf k \cdot \mathbf r_{12}}{2} \Big) {|\xi_{k}(t)|^2} , \\
 	  \gamma_2 (t) = &8 \sum_{\mathbf k} \sin^2\Big (\frac{\mathbf k \cdot \mathbf r_{12}}{2} \Big) {|\xi_{k}(t)|^2} .
 \end{align}

\subsection{The frequency shift} 

Using Eqs. \eqref{Axi}  and \eqref{AG12}, one can recast \eqref{AD11a}  as,
\begin{align}
 {\delta(t)}=&-\frac{2\lambda^2} {V }\sum_{\mathbf k}\bigg( \frac{k_z^2 \cos (\mathbf k\cdot \mathbf r_{12})(\omega_k t
	 -\sin(\omega_kt))}{\omega_k^3}\nonumber \\
	&+  \frac{2k_z^2 \sin (\mathbf k\cdot \mathbf r_{12})\sin^2(\omega_k t/2)}{\omega_k^3}\bigg )
	 \label{AD9a}.
	 \end{align}
 To proceed further, we replace a sum by an integral: $(1/V )\sum_{\mathbf k} \rightarrow (1/(2\pi)^3)\int d^3k$, and write the total phase as $\delta= \delta_1 + \delta_2$, where,
\begin{align}\label{G1}
\delta_1=&-\frac{\lambda^2}{4 \pi^3}\int\frac{ d^3k k_z^2 \cos (\mathbf k\cdot \mathbf r_{12})(\omega_k t
	 -\sin(\omega_kt))}{\omega_k^3}, \\
\delta_2 =& -\frac{\lambda^2}{2 \pi^3}\int \frac{d^3k k_z^2 \sin (\mathbf k\cdot \mathbf r_{12})\sin^2(\omega_kt/2)}{\omega_k^3}.	 
 \label{G2}
 \end{align}
	
We assume that spins are located at the points with the position vectors, 
$\mathbf r_1$ and $\mathbf r_2$. Without loss of generality, one can choose the orientation of the coordinate system in such a way that $ \mathbf r_{12}=L(\sin \theta, 0, \cos \theta)$, where $L= |\mathbf r_{12}|\equiv r_{12}$. Using the spherical coordinates $(k,\vartheta,\varphi)$, one can recast Eqs. \eqref{G1}, \eqref{G2} as,
\begin{align}\label{G1a}
\delta_1=&-\frac{\lambda^2}{2 \pi^2}\int^\infty_0\frac{ dk k^4 I_1(kL)(\omega_k t
	 -\sin(\omega_kt))}{\omega_k^3}, \\
\delta_2 =& -\frac{\lambda^2}{ \pi^2}\int^\infty_0 \frac{dk k^4 I_2(kL)\sin^2(\omega_kt/2)}{\omega_k^3}.	 
 \label{G2a}
 \end{align}
 where, 
  \begin{align}
 	I_1(kL) =\frac{1}{2\pi} \int_0^{2\pi} d\varphi\int_0^{\pi} d\vartheta \cos^2\vartheta \sin\vartheta\cos (\mathbf k\cdot \mathbf r_{12}), \nonumber \\
 	 I_2(kL) = \frac{1}{2\pi} \int_0^{2\pi} d\varphi\int_0^{\pi} d\vartheta \cos^2\vartheta \sin\vartheta\sin (\mathbf k\cdot \mathbf r_{12}).
 \end{align}

 The computation of the integrals $I_{1,2}$ yields (see Appendix B),
\begin{align}
I_1(x)=	& \frac{2}{x^3}\big ( x^2\sin x +3x\cos x  - 3\sin x\big ) \cos^2\theta_0 \nonumber \\
&+\frac{2}{x^3}\big ( \sin x -x\cos x \big ), \\
I_2(x) =&0,
\label{SI1}	
\end{align}
where a new dimensionless variable, $x = kL$, is introduced.
Since $I_2 =0$, we conclude that $\delta_2 =0$.

Then, Eq. \eqref{AD9a}, determining the total phase, can be rewritten as,
\begin{align}
\delta=-\frac{\lambda^2}{2\pi^2 L^3}\int^\infty_0\frac{ dx x^4 I_1(x)(\omega t- L\sin(\omega t/L)}{\omega^3} ,
 \label{G5}
 \end{align}
 where $\omega = \sqrt{x^2 + (mL)^2}$. 
 
 In what follows, we assume that $mL\ll 1$. Then, one can recast \eqref{G5} as,
 \begin{align}
\delta=\frac{\lambda^2 S_0(t)}{2\pi^2 L^3} + {\mathcal O} \big ((mL)^2\big ).
 \label{G5f}
 \end{align}
 where 
  \begin{align}
S_0(t)=\int^\infty_0{ dx x I_1(x)\big (L\sin(x t/L) -x t \big)} .
 \end{align}
Performing the integration, we obtain  (for detail see Appendix B),
\begin{align}
\delta (t)=\left \{\begin{array}{ll}
	0, & {\rm if} \,t < L, \\
	\displaystyle \frac{\lambda^2 t}{\pi}\cdot\frac{(3\cos^2\theta -1)}{ L^3},& {\rm if}\, t > L. 
\end{array}
\right .
	 \label{G6}
\end{align}

\subsection {Decoherence}

Here we limit ourselves by considering only the spin decoherence produced by the vacuum field.
By replacing a sum by an integral, we have,
\begin{align}
	 \gamma(t)=&\frac{\lambda^2} {4\pi^3 }\int \frac{d^3k 
	 k_z^2\sin^2(\omega_kt/2)}{\omega_k^3} , \\
	  \gamma_1(t)=&\frac{\lambda^2} {\pi^3 }\int \frac{d^3k 
	 k_z^2 \cos^2\Big (\frac{\mathbf k \cdot \mathbf r_{12}}{2} \Big)\sin^2(\omega_kt/2)}{\omega_k^3}, \\ 
	  \gamma_2(t)=&\frac{\lambda^2} {\pi^3 }\int \frac{d^3k 
	 k_z^2 \sin^2\Big (\frac{\mathbf k \cdot \mathbf r_{12}}{2} \Big)\sin^2(\omega_kt/2)}{\omega_k^3}.
	 \end{align}

In what follows, we restrict ourselves by consideration only $\gamma(t)$.
Performing integration over angle variables we obtain, 
\begin{align}
 {\gamma(t)}=&\frac{\lambda^2} {3\pi^2}\int_0^{\infty} \frac{dk 
	 k^4\sin^2(\omega_kt/2)}{\omega_k^3}.
\end{align}

The asymptotic value of $\gamma(t)$, while $t \rightarrow \infty$, is given by,
\begin{align}
	\gamma (t) \rightarrow \gamma_0 	=\frac{\lambda^2} {6\pi^2}\int_0^{\infty} \frac{dk 
	 k^4}{\omega_k^3}.
	 \label{AG1}
\end{align}
This integral is diverging as $k \rightarrow \infty$. There are different approaches to deal with this ultra-violet (UV) catastrophe, which occurs in many domains of the field theory. One of them is to introduce in (\ref {AG1}) a cutoff, $k_c$. In this approach, the question on the value of $k_c$ usually arises. 

To regularize the divergent integral (\ref {AG1}) we use the dimensional regularization, following the procedure described in \cite{EZ}. For this purpose,  consider the auxiliary integral,
\begin{align}
\Sigma_\epsilon	= \frac{2 \pi^{\frac{3}{2}- \epsilon}}{\Gamma\left(\frac{3}{2}- \epsilon\right)} \int_0^{\infty} \frac{ 
	 x^{4- 2\epsilon} dx}{(x^2+ 1)^{3/2}}.
\end{align}  
 In order to calculate this integral we will use the formula:
\begin{align}
	\int_0^{\infty} \frac{ 
	 r^\alpha dr}{(a + br^\beta)^\gamma} = \Big(\frac{a}{b}\Big)^{\frac{\alpha +1}{\beta}} \frac{\Gamma \Big(\frac{\alpha +1}{\beta}\Big)\Gamma \Big(\gamma -\frac{\alpha +1}{\beta}\Big)}{a^\gamma \beta\Gamma(\gamma)}.
	 \label{LI}
\end{align}
The computation yields,
 \begin{align}
\Sigma_\epsilon	=\frac{  \pi^{\frac{3}{2}- \epsilon} \Gamma \Big(\frac{5}{2} -\epsilon\Big)\Gamma (\epsilon -1)}{\Gamma\left(\frac{3}{2}- \epsilon\right)\Gamma \Big(\frac{3}{2} \Big)}.
\end{align}
Using the identities $\Gamma (z+1) = z \Gamma (z)$ and $\pi^{- \epsilon}  = e^{-\epsilon \ln \pi}$,  we get
\begin{align}
\Sigma_\epsilon	=2\pi  e^{-\epsilon \ln \pi}\Big(\frac{3}{2}- \epsilon\Big)\Gamma (\epsilon -1).
\end{align}

For $\epsilon \ll 1$, we have $e^{-\epsilon \ln \pi}= 1- \epsilon\ln\pi +\mathcal  O(\epsilon^2)  $.  To proceed further, we use the Laurent series expansion in a neighborhood of the pole $z =-1$:
\begin{equation}
\Gamma (z)=\Gamma (\epsilon-1)=-\frac{1}{\epsilon} +\gamma -1+\mathcal  O\left(\epsilon\right), 
\end{equation}
where $\gamma = 0.57722$ is the Euler constant. After some algebra we obtain,
\begin{align}
\Sigma_\epsilon	= 3\pi\Big(-\frac{1}{\epsilon}+ \gamma -\frac{1 }{3}+ \frac{2 }{3}\ln \pi+\mathcal  O\left(\epsilon\right) \Big ).
\end{align}
We truncate the singular term $1/\epsilon$ of the function $\Sigma_\epsilon	$ at the point $\epsilon = 0$ and obtain the regularized expression,
 \begin{align}
 	[\Sigma]_{reg} = \lim_{\epsilon \rightarrow 0}[\Sigma_\epsilon]_{reg} =\pi(3\gamma -1+  2\ln \pi).
 	\label{S2}
 \end{align}

Returning to Eq.\eqref{G1}, we rewrite it as,
\begin{align}
 \gamma_0 	=\frac{\lambda^2m^2} {6\pi^2}\int_0^{\infty} \frac{ 
	 x^4 dx}{(x^2+ 1)^{3/2}},
\end{align}
where $x=k/m$. This can be recast as follows:
\begin{align}
 \gamma_0 	=\frac{\lambda^2m^2} {24\pi^3} \lim_{\epsilon \rightarrow 0}\Sigma_\epsilon.
\end{align}
Employing \eqref{S2}, we obtain,
\begin{align}
	\gamma_0\rightarrow \gamma^{reg}_0=	\frac{\lambda^2 m^2} {24\pi^2}(3\gamma -1+  2\ln \pi).
\end{align}
Since for DM the product $\lambda m \ll 1$,  we obtain $\gamma^{reg}_0\ll 1$. As a result, we have $e^{-\gamma^{reg}_0 }\approx1$. Thus, the partial phase decoherence takes place for $\rho_{01}(t)$. Similar conclusion can be made for other off-diagonal elements of the reduced spin density matrix. Note that the ``dimensional regularization procedure", used here for decoherence rate, requires additional justification, which is not a subject of this paper.

\section{Appendix B: Calculation of some useful integrals}

Here we calculate the typical integrals emergent in the main text of the paper. We start with the integrals:
\begin{align}\label{B1}
 	I_1=&\frac{1}{2\pi} \int_0^{2\pi} d\varphi\int_0^{\pi} d\vartheta \cos^2\vartheta 
 	\sin\vartheta\cos (\mathbf k\cdot \mathbf r_{12}),  \\
 	 I_2 =& \frac{1}{2\pi} \int_0^{2\pi} d\varphi\int_0^{\pi} d\vartheta \cos^2\vartheta 
 	 \sin\vartheta\sin (\mathbf k\cdot \mathbf r_{12}).
 	 \label{B2}
 \end{align}
Here $\mathbf k =k (\sin \vartheta \cos \varphi, \sin \vartheta \sin \varphi, \cos 
\vartheta)$. Without loss of generality, we can choose the orientation of the coordinate 
system in such a way that $ \mathbf r_{12}=L(\sin \theta, 0, \cos \theta)$, where, as before, 
$L= |\mathbf r_{12}|\equiv r_{12}$. Before proceeding further, it is convenient to introduce new variables: $p = kL \sin \theta$ and $q = kL \cos \theta$. We obtain,
\begin{align}
\mathbf k\cdot \mathbf r_{12} = p \sin \vartheta \cos \varphi +  q \cos\vartheta .
\label{Kr}
\end{align}

First,  consider the integral $I_1$ written as,
\begin{align}
I_1 = \int_0^{\pi} f_1(\vartheta)\cos^2\vartheta \sin\vartheta   d\vartheta ,
\label{B3e}
\end{align}
where
\begin{align}
f_1(\vartheta)	=&\frac{1}{2\pi} \int_0^{2\pi} \cos(p\sin \vartheta \cos \varphi +  
pq\cos\vartheta ) d\varphi \nonumber \\
&=\cos(  q \cos\vartheta ) J_0(p\sin \vartheta).
\end{align}
Here $J_\nu (x)$ denotes the Bessel function. Introducing a new variable, $z= 
\sin\varphi$, one can recast \eqref{B3e} as,
\begin{align}
I_1 = 2\Re\int_0^{1} z\sqrt{1-z^2} e^{iq\sqrt{1-z^2}} J_0(pz)  dz  .
\label{B7}
\end{align}
Performing the integration, we obtain \cite{PRU}
\begin{align}
I_1(kL) = -2\frac{\partial^2}{\partial q^2} \bigg( \frac{\sin\sqrt{p^2 + q^2}}{\sqrt{p^2 + q^2}} \bigg ).
\label{B8}
\end{align}
This yields
\begin{align}
I_1(x)=	& \frac{2}{x^3}\big ( x^2\sin x +3x\cos x  - 3\sin x\big ) \cos^2\theta 
\nonumber \\
&+\frac{2}{x^3}\big ( \sin x -x\cos x \big ).
\label{B9}	
\end{align}

Next,  consider the integral $I_2$. Using \eqref{Kr}, one can recast \eqref{B2} as,
\begin{align}
I_2 =2\int_0^{\pi/2} f_2(\vartheta)\cos^2\vartheta \sin\vartheta \cos(q\cos 
\vartheta))  d\vartheta ,
\end{align}
where
\begin{align}
f_2(\vartheta)	=\frac{1}{2\pi} \int_0^{2\pi} \sin(p\sin \vartheta \cos \varphi ) 
d\varphi.
\end{align}
As one can easily see, $f_2(\vartheta) =0$, and thus the integral $I_2=0$.

Now, consider the integral,
\begin{align}  \label{BS1}
S_0(t)=\int^\infty_0 I_1(x)(L\sin(xt/L) - x t)xdx ,
 \end{align} 
 where $L > 0$. To evaluate this integral, we use the identities $(n \geq 0)$:
\begin{align}\label{AI1}
	\int_0^\infty \tau^n\cos{(\kappa \tau)}d\tau = \left \{\begin{array}{ll}
	(-1)^{n/2}\pi \delta^{(n)}(\kappa), & \quad n \quad \rm even\\
	& \\
	(-1)^{(n+1)/2}\displaystyle\frac{n!}{\kappa^{n+1}},  & \quad n \quad \rm  odd
	\end{array}
	\right .
	\end{align}
\begin{align}
\int_0^\infty \tau^n\sin{(\kappa \tau)} d\tau = \left \{\begin{array}{ll}
	(-1)^{n/2}\displaystyle\frac{n!}{\kappa^{n+1}}, & \quad n \quad \rm even\\
		& \\
	(-1)^{(n+1)/2}  \pi \delta^{(n)}(\kappa), & \quad n \quad \rm  odd
	\end{array}
	\right .
	\label{AI2}
\end{align}
where $\delta^{(n)}$ denotes the $n^{th}$ derivative of the Dirac delta function.

Taking into account that,
\begin{align}
	\int_0^\infty \frac{\sin x}{x}dx = \frac{\pi}{2},
\end{align}
after some algebra we obtain,
\begin{align}
S_0 (t)= \pi t(3 \cos^2 \theta -1)   + I_0,
 \label{B3}
 \end{align} 
 where 
 \begin{align}
 	I_0 =  L\int^\infty_0 I_1(x)\sin(x t/L)xdx. 
 \end{align}
The computation yields,	
\begin{align}
&I_0 = L\int^\infty_0 I_1(x)\sin(x t/L)xdx =\pi L \delta( t/L -1)\nonumber \\
& - \pi L\delta ( t/L +1) + \frac {\pi  t}{2}(3 \cos^2 \theta -1)\big 
(H( t/L-1)-1\big),
\label{B3a}
 \end{align} 
where $H(x)$ denotes the Heaviside function. Finally, we obtain
\begin{align}
	S_0(t)  =\left \{\begin{array}{ll}
	0, & \quad  t < L\\
2\pi t(3 \cos^2 \theta -1), & \quad  t > L
	\end{array}
	\right .
\end{align}

  \end{document}